\documentclass[reprint, superscriptaddress, secnumarabic, amsmath, amssymb, nobibnotes, aps, prb]{revtex4-2}

\usepackage{graphicx}
\usepackage{caption}
\usepackage{epstopdf}
\usepackage[T1]{fontenc}
\usepackage{amsbsy}
\usepackage{gensymb}
\usepackage{amsmath}
\usepackage{amssymb}
\usepackage{bbm}
\usepackage{braket}
\usepackage{xcolor}
\usepackage{multirow}
\allowdisplaybreaks
\usepackage{graphicx}
\usepackage[hidelinks]{hyperref}  
\usepackage[normalem]{ulem}
\usepackage{orcidlink}

\renewcommand{\approx}{\simeq}

\begin{document}

	\title{\textrm{Fractionalized Magnetization Plateaus in the Shastry-Sutherland Lattice Material Er$_2$Be$_2$GeO$_7$}}
\author{M. Pula\,\orcidlink{0000-0002-4567-5402}}
\email[]{pulam@mcmaster.ca}
\affiliation{Department of Physics and Astronomy, McMaster University, Hamilton, Ontario L8S 4M1, Canada}
\author{S. Sharma\,\orcidlink{0000-0002-4710-9615}}
\affiliation{Department of Physics and Astronomy, McMaster University, Hamilton, Ontario L8S 4M1, Canada}
\author{J. Gautreau}
\affiliation{Department of Physics and Astronomy, McMaster University, Hamilton, Ontario L8S 4M1, Canada}
\author{Sajilesh K. P.}
\affiliation{Physics Department, Technion-Israel Institute of Technology, Haifa 32000, Israel}
\author{A. Kanigel}
\affiliation{Physics Department, Technion-Israel Institute of Technology, Haifa 32000, Israel}
\author{C. R. dela Cruz}
\affiliation{Neutron Scattering Division, Oak Ridge National Laboratory, Oak Ridge, Tennessee 37831, USA}
\author{T. N. Dolling\,\orcidlink{0009-0007-9324-1485}}
\affiliation{School of Chemistry, University of Birmingham, Edgbaston, Birmingham B15 2TT, UK}
\author{L. Clark\,\orcidlink{0000-0002-6223-3622}}
\affiliation{School of Chemistry, University of Birmingham, Edgbaston, Birmingham B15 2TT, UK}
\author{G.~M.~Luke\,\orcidlink{0000-0003-4762-1173}}
\email[]{luke@mcmaster.ca}
\affiliation{Department of Physics and Astronomy, McMaster University, Hamilton, Ontario L8S 4M1, Canada}
\affiliation{TRIUMF, Vancouver, British Columbia V6T 2A3, Canada}	
	\date{\today}
	\begin{abstract}
		\begin{flushleft}
		\end{flushleft}
  The experimental study of magnetism on the Shastry-Sutherland lattice has been ongoing for more than two decades, following the discovery of the first Shastry-Sutherland lattice materials SrCu$_2$(BO$_3$)$_2$. However, the study of Shastry-Sutherland systems is often complicated by the requirements of high magnetic fields ($>$~20~T SrCu$_2$(BO$_3$)$_2$) or the presence of itinerate electrons (e.g. REB$_4$). In this paper, we present the magnetic properties of the Shastry-Sutherland lattice material Er$_2$Be$_2$GeO$_7$. Like SrCu$_2$(BO$_3$)$_2$, Er$_2$Be$_2$GeO$_7$ exhibits fractionalized magnetization plateaus. Unlike SrCu$_2$(BO$_3$)$_2$, Er$_2$Be$_2$GeO$_7$ exhibits long-range order below $\sim1~$K, and the plateaus are accessible using commercial laboratory equipment, occurring for fields <~1~T. The fractions of magnetization present are closest to $\frac{1}{4}$ and $\frac{1}{2}$ of the full powder moment; we show that the $\frac{1}{4}$ magnetization plateau in Er$_2$Be$_2$GeO$_7$ has a classical analog, well represented by the magnetic structure (canted antiferromagnetic) observed in powder neutron diffraction. The lack of itinerate electrons, chemical disorder, and the low fields required to access the fractionalized magnetization plateaus promises Er$_2$Be$_2$GeO$_7$ to be a prime candidate for the study of frustrated magnetism on the Shastry-Sutherland lattice.

	\end{abstract}
	\maketitle

\section{Introduction}

Magnetic frustration occurs when competing magnetic interactions coexist within a system. Unlike systems with co-operative interactions, which facilitate magnetic order, these competing interactions tend to combat magnetic order, reducing the transition temperature for which ordering occurs or outright preventing it. The influence frustration has on the system stems from the mutual exclusivity in minimizing the energy of the competing interactions; this leads to a ground-state degeneracy that acts to impede magnetic order \cite{rau2019frustrated, lacroix2011introduction, balents2010spin}. 

Frustration that occurs inherently due to the arrangement of magnetic ions on a particular lattice is known, unsurprisingly, as geometric frustration. This type of frustration is ubiquitous in condensed matter physics, and several archetypal geometrically frustrated systems have emerged in the literature, including the pyrochlore, kagom\'e, spinel, triangular, and Shastry-Sutherland lattices. 

Geometrically frustrated magnets are known for their unconventional magnetic ground states. In the specific case of the Shastry-Sutherland lattice (SSL), a theoretical model consisting of diagonal next-nearest-neighbour ($J^{'}$) and square nearest-neighbour ($J$) antiferromagnetic exchange interactions on a square lattice, the ground state was predicted  to be one of two states: a dimer singlet or a square antiferromagnet \cite{SRIRAMSHASTRY19811069}. The exact ground state is determined by the ratio of the characteristic exchange interactions of the lattice ($J^{'}/J$). The ground state will be the square antiferromagnet when $J^{'}\gg J$ , while $J^{'}\ll J$ leads to the dimer singlet; this dimer state was determined to be the ground state of SrCu$_2$(BO$_3$)$_2$ \cite{PhysRevLett.82.3168}, whose lattice differs from but is topologically equivalent to the ``toy'' model imagined by Shastry and Sutherland.

SrCu$_2$(BO$_3$)$_2$ exhibits a spin gap, with an energy separation between the spin-singlet ground state and the excited triplet state, of $\Delta$~=~30~K and was the first two-dimensional spin system to exhibit fractional magnetization plateaus \cite{PhysRevLett.82.3168}. Despite the initially predicted dichotomy of possible ground states in the Shastry-Sutherland model, subsequent theoretical investigations concluded that there is an additional plaquette phase in the intermediate $J^{'}/J$ region \cite{koga2000quantum}; indeed, experimentation performed under pressure (acting to alter the $J^{'}/J$ ratio) confirmed this plaquette state occurs in SrCu$_2$(BO$_3$)$_2$ \cite{guo2020quantum,jimenez2021quantum}. Most recently, a quantum spin liquid phase has also been predicted \cite{Wang_2022, PhysRevB.105.L041115, PhysRevB.105.L060409}, which may have been realized in another SSL system, the rare-earth melilite-like Yb$_2$Be$_2$GeO$_7$ \cite{pula2024candidate}.

\begin{figure}
    \centering
    \includegraphics[width=\columnwidth]{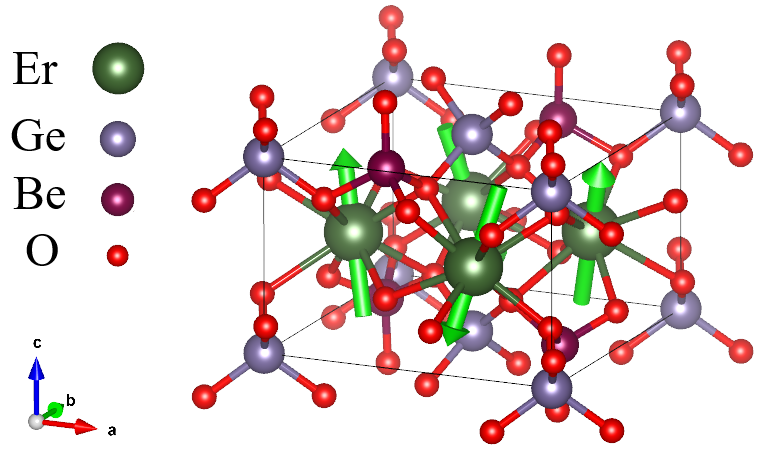}
    \caption{Er$_2$Be$_2$GeO$_7$ lattice (space group 113) with spins along the local z-axis, as predicted via PCCEF calculations. Generated via VESTA \cite{momma2011vesta}.}
    \label{fig:Re2Be2GeO7-lattice}
\end{figure}

 The rare-earth melilites  were first synthesized by Ochi {\em et al.} in 1982 \cite{OCHI1982911} and first recognized as topologically equivalent to the SSL by Ashtar and Bai \cite{doi:10.1021/acs.inorgchem.0c03131}. They are comprised of alternating layers of magnetic RE$^{3+}$ ions and non-magnetic Be$_2$GeO$_7$. In particular, the geometry of the RE$^{3+}$ ions within the RE$^{3+}$ layers is topologically equivalent to the SSL motif (see Fig. \ref{fig:Re2Be2GeO7-lattice}), and many of the lanthanides can occupy the RE site. Furthermore, rare earth melilites have been reported to lack detectable site mixing of their constituent atoms \cite{doi:10.1021/acs.inorgchem.0c03131}, so chemical disorder is not an expected complication.

 The lack of chemical disorder in conjunction with the multitudinous nature of the RE site (implying possible tunability of the single-ion anisotropy) promises a system in which to study geometric frustration on the SSL that is perhaps as diverse as the well-studied pyrochlores. Currently, little is known about the system, especially the magnetic properties. Ashtar and Bai reported magnetometry measurements down to 2~K, for which they found that all but one (the Tb-based melilite) lack magnetic order. However, recent work on Yb-based melilite \cite{pula2024candidate} and an analogous SSL material, Er$_2$Be$_2$SiO$_7$ \cite{brassington2024magnetic}, has shown that the salient magnetic property changes occur below 2~K, a temperature region for which studies have not yet been reported. In this paper, we report our findings on the magnetic properties of one of these rare-earth melilites, Er$_2$Be$_2$GeO$_7$,finding an antiferromagnetic ground state below about $T=1$~K, which  hosts fractionalized magnetization plateaus. 

\begin{figure}
    \centering
    \includegraphics[width=\columnwidth]{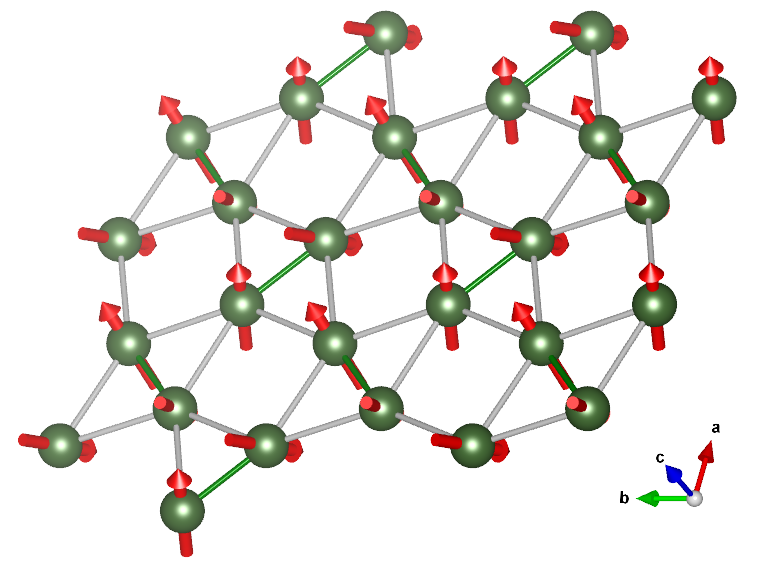}
    \caption{Intraplanar arrangement of Er ions, which is topologically equivalent to the Shastry-Sutherland lattice when the nearest-neighbour (green bond) and next-nearest neighbour (grey bonds) magnetic interactions are both antiferromagnetic. Red arrows represent the magnetic space group 18.19, with parent crystallographic space group 113. Here, M = [1.39(5), 1.01(6), -4.18(2)] $\mu_B$/Er$^{3+}$. Generated via VESTA \cite{momma2011vesta}.}
    \label{fig:intraplanarspins}
\end{figure}

\section{Methods}

Powder Er$_2$Be$_2$GeO$_7$ was synthesized by a solid state reaction, in which a stoichiometric mixture of Er$_2$O$_3$ (4N), BeO (4N) and GeO$_2$ (4N) was first ground in an argon glovebox, then subsequently heated to 1350~\degree~C. This process was repeated multiple times (on average, 4) and is based on the work of Y. Ochi et al.\cite{OCHI1982911}, and Y. Bai and M. Ashtar\cite{doi:10.1021/acs.inorgchem.0c03131}.

\begin{figure*}
    \centering
    \includegraphics[width=\textwidth]{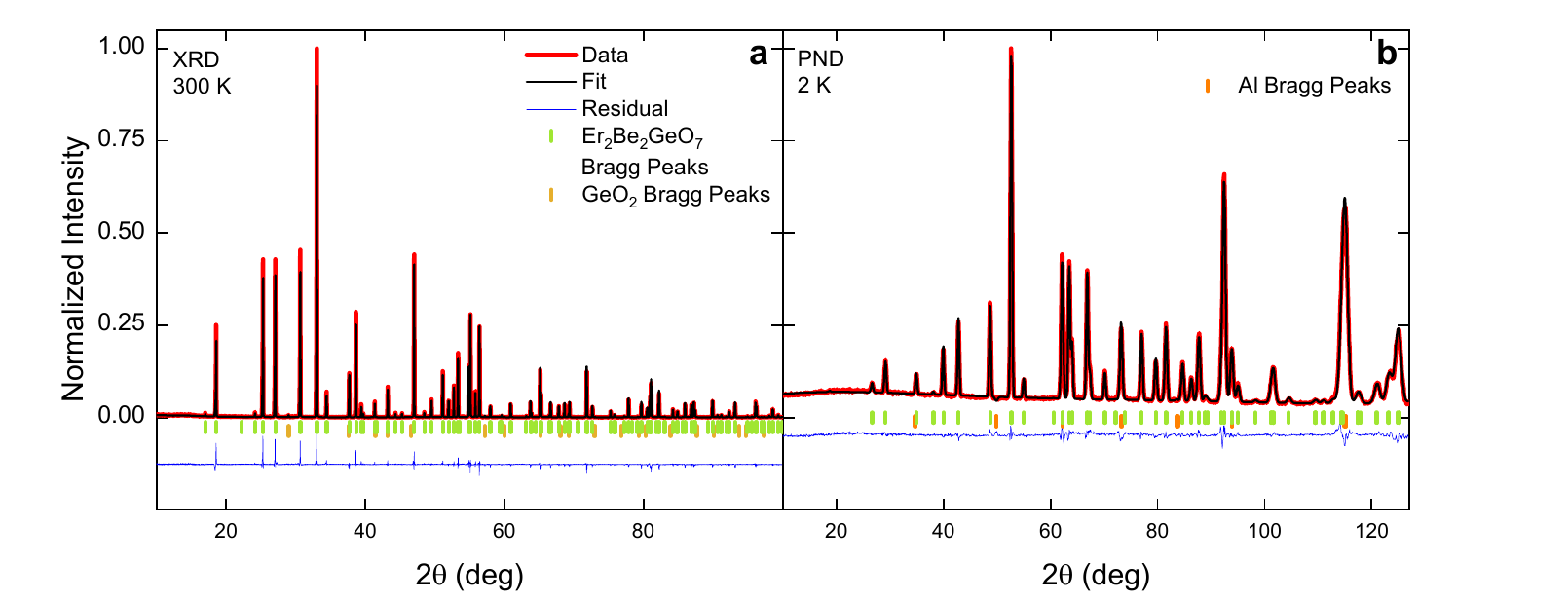}
    \caption{\textbf{a}: XRD of Er$_2$Be$_2$GeO$_7$ at 300~K. A small GeO$_2$ impurity is visible around 30 degrees. This was included in the refinement. The Bragg peaks of the main phase Er$_2$Be$_2$GeO$_7$ are indicated by the green ticks. \textbf{b}: PND of Er$_2$Be$_2$GeO$_7$ at 2~K. The legends of \textbf{a} and \textbf{b} are shared. The contribution of the aluminum sample can is indicated by the orange ticks.}
    \label{fig:Er-Ho-Structrual}
\end{figure*}

The structure of the sample was determined by X-ray diffraction (XRD) conducted on a Panalytical X'pert Pro diffractometer. Bulk magnetic measurements were performed on a Quantum Design MPMS XL magnetometer with an iQuantum helium-3 insert and a Quantum Design MPMS3 magnetometer with a helium-3 insert. Powder neutron diffraction (PND) measurements were collected at Oak Ridge National Laboratory on the High Flux Isotope Reactor's POWDER diffractometer. The loose powders were sealed in an aluminum sample can with $\approx$ 1 bar of helium-4 gas. Specific heat capacity measurements utilized the two-tau relaxation method on a Quantum Design PPMS equipped with a dilution refrigerator. La$_2$Be$_2$GeO$_7$ was used as a non-magnetic analog to remove the contribution of phonons to the specific heat. 

The point charge crystal electric field (PCCEF) calculations employed the PyCrystalField python package\cite{scheie2021pycrystalfield}. The structural details (i.e., atomic positions and lattice parameters) used in these calculations were determined from the XRD Rietveld refinement (see Fig. \ref{Tab:Er-Ho-Rietveld}). The eight nearest-neighbor oxygen ligands were considered in the calculation.

\section{results}

\begin{table}[b!]
    \centering
    \resizebox{\columnwidth}{!}{
    \begin{tabular}{|c|c|c|c|c|}
    \hline
        Range &  C ($\frac{cm^3}{mol K}$) & $\theta$ (K) & $\mu_{eff}$ ($\mu_B$) & $\mu_{J}$ ($\mu_B$) \\
         \hline
        300K $\rightarrow$ 33K & 22.277(4)  & -6.21(4) & 9.411(2) & 9.581 \\
        20K $\rightarrow$ 5K & 19.77(7)  & -2.89(4) & 8.87(3) & 9.58 \\
        \hline
        
    \hline    
    \end{tabular}}
    \caption{Summary of the Curie-Weiss fitting parameters.}
    \label{tab:CurieWeissPara}
\end{table}

X-ray diffraction (XRD) and powder neutron diffraction (PND) were used to confirm the crystallographic structure (see Fig. \ref{fig:Re2Be2GeO7-lattice} for a visualization) of Er$_2$Be$_2$GeO$_7$, which is the tetragonal space group P$\overline{\mbox{4}}$2$_1$m (113). The XRD pattern shows the presence of a slight impurity phase, evidenced by an additional Bragg peak around 29\degree. This is attributed to unreacted GeO$_2$, whose highest intensity peak occurs around 29\degree~\cite{baur1956verfeinerung}. The absence of additional Bragg peaks indicates that the compositional fraction of GeO$_2$ is relatively small compared to the main phase, and GeO$_2$ is known to be non-magnetic \cite{LandoltBornstein2007} and to have no influence on the magnetic properties of the isomorphic rare-earth melilite Yb$_2$Be$_2$GeO$_7$ \cite{pula2024candidate}.

As such, it is not expected to affect the measurements conducted to any considerable degree. Rietveld refinement was used to extract the structural details from the patterns. The lattice parameters for Er$_2$Be$_2$GeO$_7$ are found from XRD at 300~K to be a~=~b~=~7.35599(2)~\AA \space and c~=~4.76698(1)~\AA; PND at 2~K results in lattice parameters of a~=~b~=~7.3817(1)~\AA \space and c~=~4.7813(1)~\AA. A summary of the structural parameters derived via the Rietveld refinement can be found in Table \ref{Tab:Er-Ho-Rietveld}, while the diffraction patterns can be found in Fig. \ref{fig:Er-Ho-Structrual}. \newline

\begin{figure*}
    \centering
    \includegraphics[width=\textwidth]{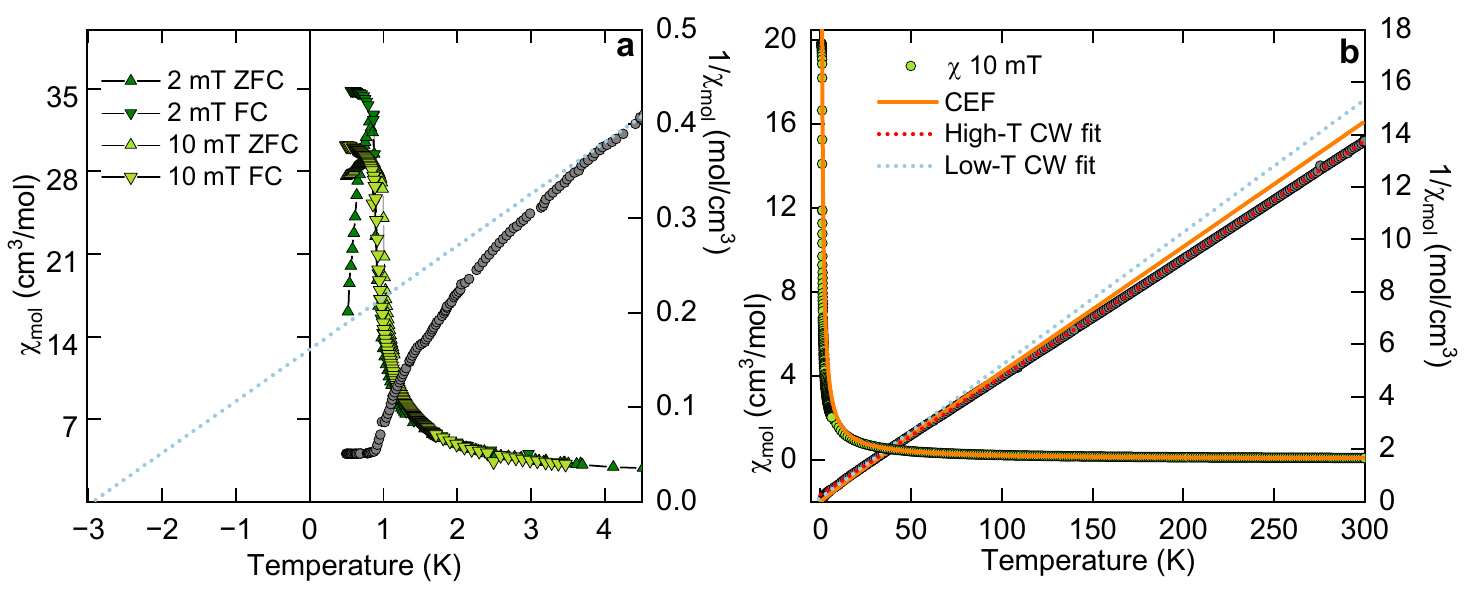}
    \caption{\textbf{a} The low-temperature magnetic susceptibility of Er$_2$Be$_2$GeO$_7$. The blue dotted line is the low-temperature Curie-Weiss fit. The displayed inverse susceptibility is combined ZFC+FC at 10 mT. A weak bifurcation can be seen in a small field of 2 mT, which is largely absent by 10 mT. \textbf{b}: High-temperature magnetic susceptibility of Er$_2$Be$_2$GeO$_7$. The susceptibility predicted by the PCCEF model is shown as a solid orange line. The blue dotted line corresponds to the Curie-Weiss law at low temperatures, while the red dotted line to high temperatures. Parameters for the Curie-Weiss fittings of each panel can be found in Table \ref{tab:CurieWeissPara}.}
    \label{fig:MagSusc}
\end{figure*}

The magnetic susceptibility of Er$_2$Be$_2$GeO$_7$ is typical of systems with crystal electric fields (CEFs). Two distinct high-and low-temperature slopes are observed in the inverse susceptibility (see Figs. \ref{fig:MagSusc}.b and \ref{fig:MagSusc}.d), indicative of changes in the occupation of CEF levels as the thermal energy changes. This slope change occurs around $\approx$~20~K. The parameters characterized by these two slopes are tabulated in Table~\ref{tab:CurieWeissPara}.

1/$\chi$ remains linear down to $\sim$~5~K in Er$_2$Be$_2$GeO$_7$(see Fig. \ref{fig:MagSusc}.a). Below these temperatures, a continuous deviation from linearity occurs, where the slope of $1/\chi$ increases with decreasing temperature; this is likely associated with an impending phase transition, for which the Curie-Weiss law does not account. In fact, the low-temperature susceptibility reveals a transition around 1~K. An abrupt change in the temperature dependence of susceptibility develops, along with a bifurcation between zero-field cooled (ZFC) and field-cooled (FC) measurements, in fields on the order of 1~mT. This irreversibility is largely eliminated in fields as low as 10~mT. Specific heat capacity and PND measurements, shown in the following, confirm that long-range magnetic ordering occurs in Er$_2$Be$_2$GeO$_7$. \newline

\begin{figure*}
    \centering
    \includegraphics[width=\textwidth]{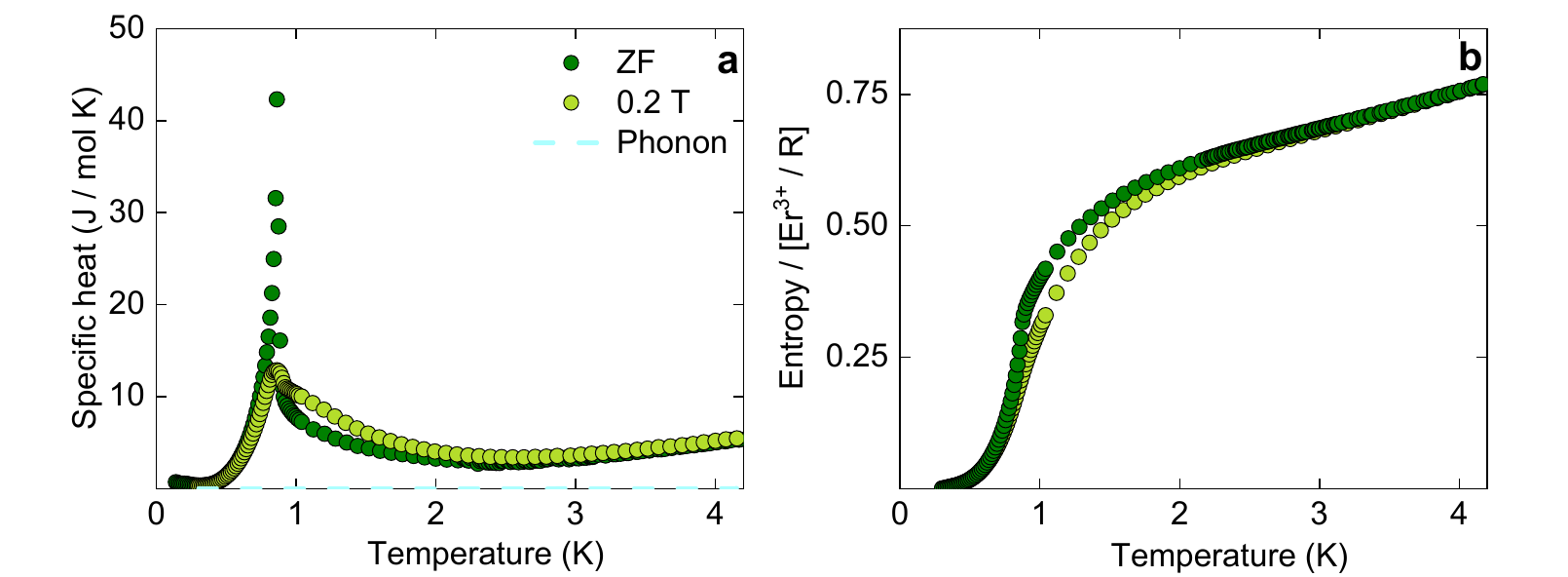}
    \caption{The magnetic specific heat capacity \textbf{a} and entropy \textbf{b} of Er$_2$Be$_2$GeO$_7$. In \textbf{a}, a clear $\lambda$-shaped peak is observed around 0.85 K, indicative of a phase transition. In \textbf{b}, the entropy can be seen to exceed R~ln(2). La$_2$Be$_2$GeO$_7$ was used as a non-magnetic analog to determine the phonon contribution (cyan dashed line), which was subtracted from the specific heat capacities of both samples. The entropy was calculated as $S=\int \frac{C_p}{T} dT$.}
    \label{fig:Ho-Er-specific-heat}
\end{figure*}

The magnetic specific heat measured in zero magnetic field shows a clear $\lambda$-like anomaly, marking the onset of magnetic order at 0.85~K. The expected entropy of an effective spin-$\frac{1}{2}$ system is Rln(2) per spin. Despite the prediction of the PCCEF calculation (see Table \ref{tab:Er-g-factor}) that Er$_2$Be$_2$GeO$_7$ is an effective spin-$\frac{1}{2}$ system, it can be seen that the entropy exceeds this expectation. This is likely an indication that at least one excited level lies at an energy much closer to the ground state than the calculated 12.26 meV. The magnetic specific heat in field (specifically, a 0.2~T field) is skewed towards higher temperatures(see Fig. \ref{fig:Ho-Er-specific-heat}.a) and shows a suppression of the cusp at the transition temperature. The entropy (see Fig. \ref{fig:Ho-Er-specific-heat}.b) near the transition is more gradual in field, but still reaches an equivalent value by $\sim$~3~K.\newline

\begin{figure*}
    \centering
    \includegraphics[width=\textwidth]{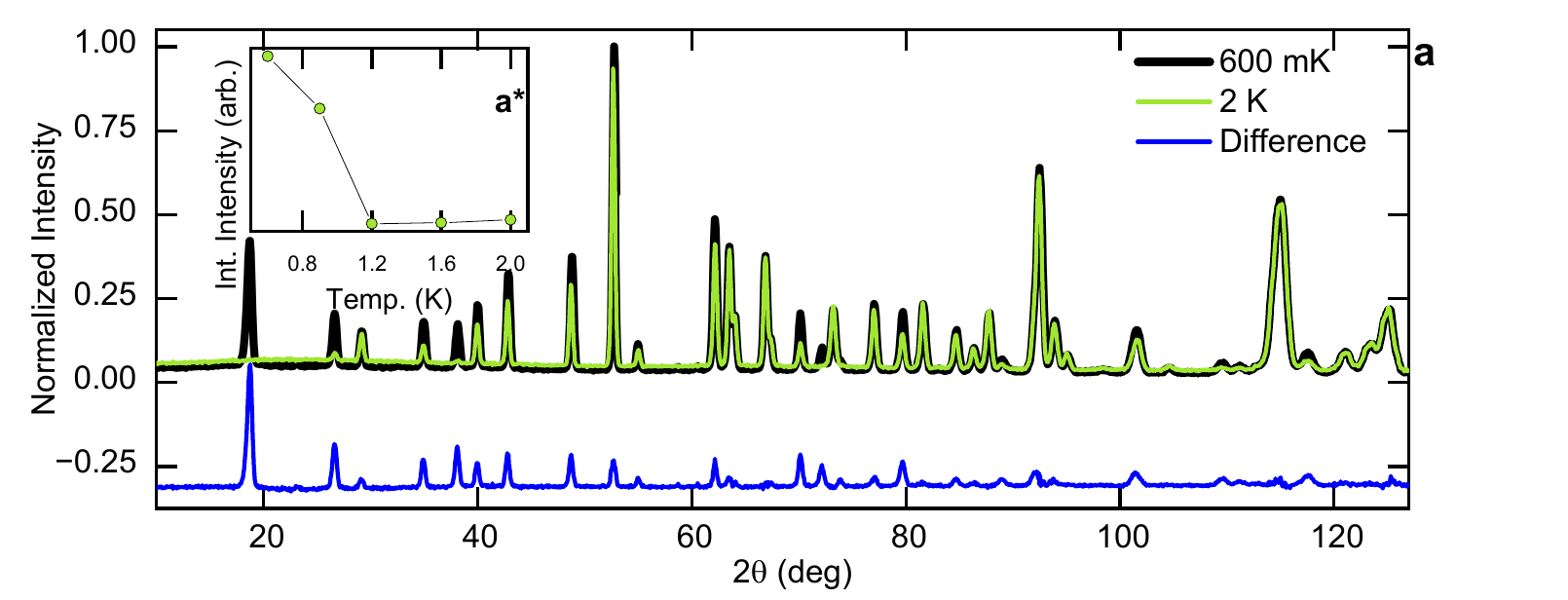}
    \caption{PND of Er$_2$Be$_2$GeO$_7$. Magnetic Bragg peaks in Er$_2$Be$_2$GeO$_7$ are isolated by subtracting the pattern above the transition temperature (600~mk) from the pattern below it (2~K). The resulting difference is the blue line. \textbf{a*}: The order parameter of Er$_2$Be$_2$GeO$_7$, calculated as the integrated intensity of the full intensity peak from 17 to 20 degrees.}
    \label{fig:isolatedMagBragg}
\end{figure*}

\begin{table*}
\begin{ruledtabular}
    \begin{tabular}{c|c|cccc|c|ccc}
    PND (2~K) & Atom & x& y& z &Occ. &Phase &a, \AA&b, \AA& c, \AA \tabularnewline
    \hline
     &Er & 0.1579(3) & 0.3421(3) & 0.5056(5) & 0.5& {Er$_2$Be$_2$GeO$_7$} &7.3817(1) &7.3817(1)& 4.7813(1) \tabularnewline
     &Be & 0.1346(2)& 0.6346(2) & 0.0472(5) &0.50 & &  & &  \tabularnewline
     &Ge & 0.0 & 0.0 & 0.0 & 0.250  &  &R$_p$ &R$_{wp}$ & $\lambda$, \AA \tabularnewline
     &O$_1$ & 0.0820(3) & 0.8285(3) & 0.2176(6) &1 & & 8.22 & 8.25& 1.486 \tabularnewline
     &O$_2$ & 0.1426(3) & 0.6426(3) & 0.7135(6) &0.50 &  &  & & \tabularnewline
     &O$_3$ & 0 & 0.5 & 0.189(1) & 0.25 &  & &  & \tabularnewline
    \hline
     XRD (300~K)&Er & 0.15864(8) & 0.34136(8) & 0.5073(3) & 0.5&Phase &a, \AA&b, \AA& c, \AA \tabularnewline
     &Be & 0.133(2)& 0.633(2) & 0.025(6) &0.50 & Er$_2$Be$_2$GeO$_7$ &7.35599(2) &7.35599(2)& 4.76698(1) \tabularnewline
     &Ge & 0.0 & 0.0 & 0.0 & 0.250 &GeO$_2$ & 4.3527(1) & 4.3527(1)& 2.8600(1) \tabularnewline
     &O$_1$ & 0.085(1) & 0.830(1) & 0.200(1) &1 & &  & & \tabularnewline 
     &O$_2$ & 0.143(1) & 0.643(1) & 0.740(2) &0.50 &  &R$_p$ &R$_{wp}$ & $\lambda$, \AA \tabularnewline
     &O$_3$ & 0 & 0.5 & 0.175(3) & 0.25 &  & 13.2 & 20.3& 1.541 \tabularnewline
     \hline \hline
    \end{tabular}
\end{ruledtabular}
\caption{Summary of the Rietveld refinement performed for both XRD and PND on Er$_2$Be$_2$GeO$_7$. Classic R$_p$ and R$_{wp}$ values are reported.}
\label{Tab:Er-Ho-Rietveld}
\end{table*}

The PND of Er$_2$Be$_2$GeO$_7$ is shown in Fig.~\ref{fig:isolatedMagBragg}.~Magnetic Bragg peaks are observed below 1~K, which are readily isolated from the lattice contribution by subtracting a high (disordered) temperature pattern (see Fig. \ref{fig:isolatedMagBragg}.a* for an order parameter).

The magnetic space group in Er$_2$Be$_2$GeO$_7$ was determined to be P2$_1$2$_1$'2', with lattice parameters a~=~7.3797(2)~\AA, b~=~7.3763(2)~\AA, and c~=~4.7784(1)~\AA. The magnetization vector at each Er site at 0.6~K is M~=~ [1.39(5), 1.01(6), -4.18(2)]~$\mu_B$. The unit cell has four Er atoms, for which the components M$_y$ and M$_z$ take both positive and negative values with equal regularity. As such, the net moment per unit cell contains a contribution from $M_x$ only, and sums to M~=~[5.6(2), 0, 0]~$\mu_B$ per unit cell, or 1.39(5) $\mu_B$/Er$^{3+}$ atom. From the magnetometry measurements, the magnetization is expected to reach $\sim$ 1.74~$\mu_B$ below H$_C^2$, which is relatively close to the moment observed in PND. Considering that there is still significant magnetic specific heat (see Fig. \ref{fig:Ho-Er-specific-heat}) around 0.6~K, it is likely that the difference is due to a lack of reaching the full moment at 0.6~K. In this structure, the nearest-neighbour spins are aligned antiferromagentically, while half of the next-nearest-neighbour are aligned ferromagnetically, and the other half antiferromagnetically (see Fig. \ref{fig:intraplanarspins}). This can be adequately described as a canted antiferromagnet, where nearest-neighbor spins align antiferromagnetically on an orthogonal lattice. For nonnegligible next-nearest-neighbor interactions, this structure has clear frustration. \newline

Point charge crystal electric field (PCCEF) calculations reveal strong Ising anisotropy (g~$\approx$~g$_z$~=~17.8) within Er$_2$Be$_2$GeO$_7$ along the local Er z-axis (see Fig. \ref{fig:Re2Be2GeO7-lattice}) of [-0.1606, 0.1606, 1]. The groundstate is a Kramer's doublet dominated by the $\pm\frac{15}{2}$ manifold, and separated from the first excited state by 12.26~meV. The predicted magnetic susceptibility somewhat agrees with the experimental susceptibility (see Fig. \ref{fig:MagSusc}. b), the difference being highlighted in the slopes of the inverse susceptibility; at high temperatures, the slope of the inverse susceptibility generated by the PCCEF is slightly overestimated, or equivalently the effective moment is underestimated. The PCCEF parameters are summarized in Table \ref{tab:Er-g-factor}.

\begin{table*}
\begin{ruledtabular}
    \begin{tabular}{c|ccccccccc}
    Er$^{3+}$&\tabularnewline
    Frame & a & b & c & G$_{ij}$ & x& y& z \tabularnewline
    \hline
    x & 1 & -1 & 0.7649&  & 0.0175 & 0 &-0.793 \tabularnewline
    y & 1 & 1 & 0 &  & 0 & 0.0314 & 0 \tabularnewline
    z & -0.1606 & 0.1606 & 1 &  & 0.00320 & 0 & 17.8 \tabularnewline
    \hline
    E (meV) & $| -\frac{15}{2}\rangle$ & $| -\frac{13}{2}\rangle$ & $|  -\frac{11}{2}\rangle$ & $| -\frac{9}{2}\rangle$ & $| -\frac{7}{2}\rangle$ & $| -\frac{5}{2}\rangle$ & $| -\frac{3}{2}\rangle$ & $| -\frac{1}{2}\rangle$ \tabularnewline
    \hline
    0.000 & -0.98 & 0.1069 & 0.1559 & -0.0217 & -0.0517 & -0.0017 & 0.0224 & 0.0029 \tabularnewline
    0.000 & 0.0 & -0.0012 & 0.0003 & 0.0018 & -0.0005 & -0.0044 & -0.0003 & 0.0103\tabularnewline
    12.263 & -0.066 & -0.8795 & 0.212 & 0.2999 & -0.0586 & -0.1889 & -0.0158 & 0.1278 & \tabularnewline
    12.263 & -0.0028 & 0.1301 & -0.0571 & -0.0327 & 0.0514 & 0.0154 & -0.0824 & -0.0258
    \tabularnewline
    19.459 & -0.003 & 0.2255 & -0.1219 & 0.1274 & 0.0401 & -0.3251 & -0.1407 & 0.4469
    \tabularnewline
    19.459 & -0.1058 & -0.1142 & -0.3247 & 0.112 & 0.398 & 0.0112 & -0.5009 & -0.1991
    \tabularnewline
    26.838 & 0.0189 & 0.189 & -0.0094 & 0.233 & -0.0904 & -0.4321 & -0.0669 & 0.3317
    \tabularnewline
    26.838 & -0.1308 & -0.1226 & -0.6404 & 0.1569 & 0.3556 & 0.0648 & 0.0461 & 0.023
    \tabularnewline
    \hline
    E (meV) & $| +\frac{1}{2}\rangle$ & $| +\frac{3}{2}\rangle$ & $|  +\frac{5}{2}\rangle$ & $| +\frac{7}{2}\rangle$ & $| +\frac{9}{2}\rangle$ & $| +\frac{11}{2}\rangle$ & $| +\frac{13}{2}\rangle$ & $| +\frac{15}{2}\rangle$ \tabularnewline
    \hline
    0.000&-0.0103 & -0.0003 & 0.0044 & -0.0005 & -0.0018 & 0.0003 & 0.0012 & 0.0\tabularnewline
    0.000& 0.0029 & -0.0224 & -0.0017 & 0.0517 & -0.0217 & -0.1559 & 0.1069 & 0.98 \tabularnewline
    12.263& 0.0258 & -0.0824 & -0.0154 & 0.0514 & 0.0327 & -0.0571 & -0.1301 & -0.0028 \tabularnewline
   12.263& 0.1278 & 0.0158 & -0.1889 & 0.0586 & 0.2999 & -0.212 & -0.8795 & 0.066 \tabularnewline
   
   19.459& 0.1991 & -0.5009 & -0.0112 & 0.398 & -0.112 & -0.3247 & 0.1142 & -0.1058 \tabularnewline
   
   19.459&0.4469 & 0.1407 & -0.3251 & -0.0401 & 0.1274 & 0.1219 & 0.2255 & 0.003 \tabularnewline

   26.838&0.023 & -0.0461 & 0.0648 & -0.3556 & 0.1569 & 0.6404 & -0.1226 & 0.1308 \tabularnewline

   26.838&-0.3317 & -0.0669 & 0.4321 & -0.0904 & -0.233 & -0.0094 & -0.189 & 0.0189 \tabularnewline
       \end{tabular}
    \end{ruledtabular}
    \caption{Details of the PCCEF, including reference frame, g-factor, and energy levels with manifold occupations. Energies above $\sim$300~K are excluded. The CEF reference frame is parameterized by fractional crystallographic lattice parameters.}
    \label{tab:Er-g-factor}
\end{table*}

\begin{figure*}
    \centering
    \includegraphics[width=\textwidth]{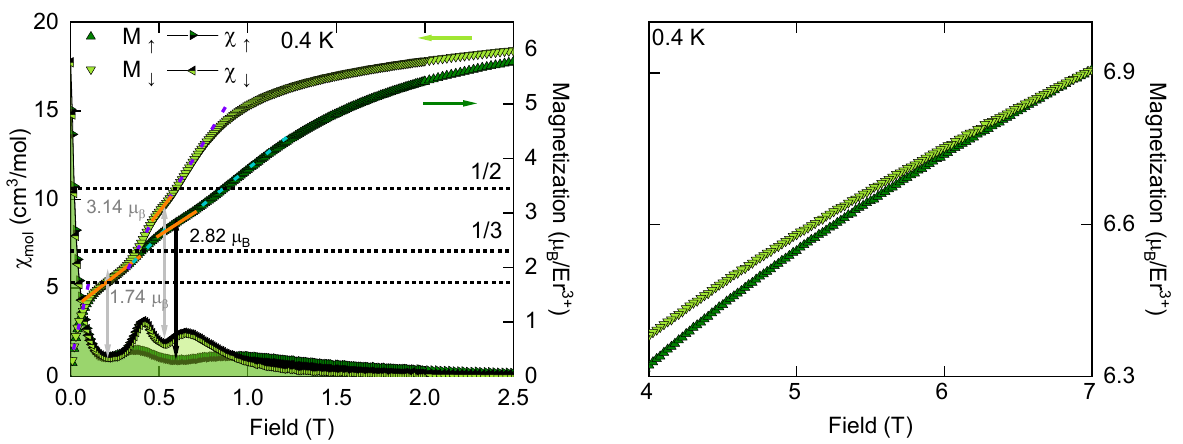}
    \caption{ \textbf{a}: Isothermal magnetization of Er$_2$Be$_2$GeO$_7$ at 0.4~K. The magnetization plateaus are visible in the field-lowering magnetization at 1.74 and 3.14~$\mu_B$/Er$^{3+}$ for fields of about 0.3 and 0.5~T, respectively. A full powder moment of 6.9~$\mu_B$/Er$^{3+}$ implies these are plateaus of $\frac{1}{4}$ and $\frac{1}{2}$ of the full powder moment. The intersection of dotted cyan (purple) lines with the solid orange lines indicates the critical fields of each plateau while field raising (lowering). \textbf{b}: Saturation magnetization of Er$_2$Be$_2$GeO$_7$ at 0.4~K.}
    \label{fig:MagPlateaus}
\end{figure*}

We measured the magnetization as a function of the applied magnetic field within the ordered states. Magnetization plateaus are indicative of fractionalized magnetization (being a rational fraction of the full moment) and themselves imply spin gaps, which prevent magnetization from continuously increasing with the applied field by presenting a finite energy barrier to magnetic excitations \cite{lacroix2011introduction}. This energy gap needs first be overcome by a critical field before the magnetization can further increase. Overcoming a plateau and entering a state with magnetization that varies continuously with applied field is synonymous with a transition from a gapped state to a gapless state; this is accompanied by a peak in the susceptibility as the magnetization goes from being constant in field to continuously increasing.

\begin{figure}
    \centering
    \includegraphics[width=\columnwidth]{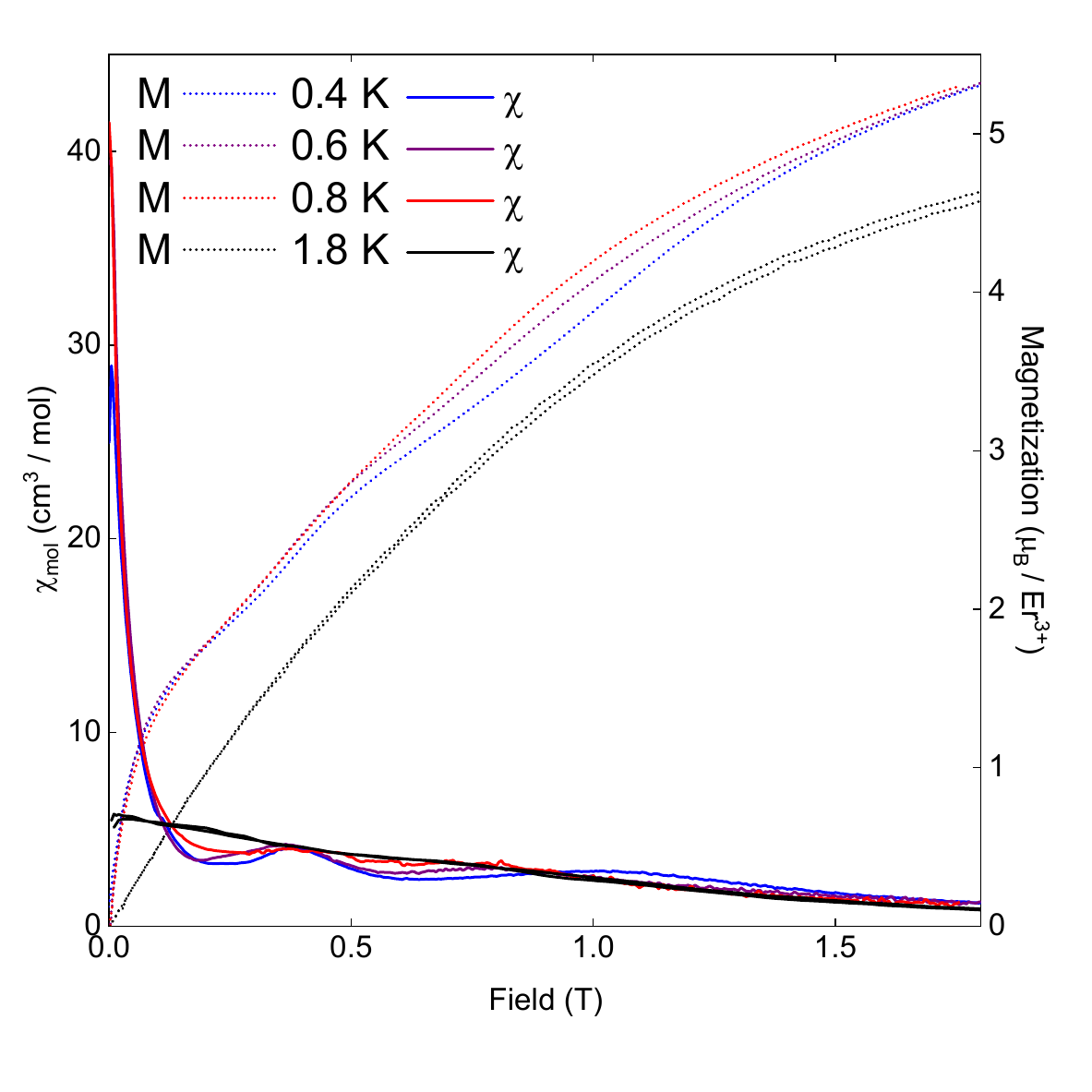}
    \caption{ \textbf{a}: $\chi(H)$ and M(H) of Er$_2$Be$_2$GeO$_7$ at four temperatures (1.8~K, 0.8~K, 0.6~K, and 0.4~K) under field-raising conditions.}
    \label{fig:Er-Ho-plats-all-temps}
\end{figure}

 In Er$_2$Be$_2$GeO$_7$, the magnetization shows plateau-like features around 0.3~T and 0.6~T (see Fig. \ref{fig:MagPlateaus}.a). 
Due to the combination of thermal fluctuations and powder averaging, the observed plateau will not reach a constant value \cite{PhysRevLett.82.3168}. The critical fields may be extracted from the magnetization as the intersection of the plateau region and the linear regions above/below it, or as the local maximums in the isothermal susceptibility. The first gapped region (plateau) is found to be bound by critical fields of about H$_c^1$ $\sim$~0.1~T and H$_c^2$ $\sim$~0.35~T. The magnetization reached upon plateauing is taken to be the local minimum in the isothermal susceptibility, which corresponds to the centre of the plateau. The first plateau reaches a moment of 1.74~$\mu_B$, corresponding well with $\frac{1}{4}$ of the full powder moment of $\sim$~6.9~$\mu_B$ at 7~T (see figure \ref{fig:MagPlateaus}.b). Interesting, in fields that exceed H$_c^2$, bifurcation occurs between the field raising and field lowering magnetizations. The field-raising plateau reaches a value of 2.82~$\mu_B$ and occurs with critical fields of H$_c^3$ $\sim$ 0.49~T and H$_c^4$ $\sim$ 0.71~T. The field-lowering plateau decreases to 3.14~$\mu_B$ and is bound by critical fields of H$_c^3$ $\sim$ 0.47~T and H$_c^4$ $\sim$ 0.58~T. The magnetization of the field-lowering plateau is roughly $\frac{1}{2}$ of the full powder moment at 7~T. The temperature dependence of the magnetization is shown in Fig. \ref{fig:Er-Ho-plats-all-temps}.

\section{Discussion}

The ground state of Er$_2$Be$_2$GeO$_7$ is clearly antiferromagnetic. The magnetic order observed is reflected in the magnetic susceptibility as a discontinuous cusp and negative Curie-Weiss temperature, the specific heat capacity as a $\lambda$-like anomaly, and PND as magnetic Bragg peaks. These features occur at the same temperature of $\sim$1~K, and the latter shows a spin structure of canted antiferromagnetic nearest-neighbor spins. The PCCEF calculations predict a Kramer's doublet effective spin-$\frac{1}{2}$ system in the $\frac{15}{2}$ manifold, separated by the first excited state by 12.26~meV; the observed magnetic entropy, however, exceeds Rln(2), implying that the first level lies much closer to the ground state. The PCCEF calculations also predict Ising single-ion anisotropy with a local z axis of [-0.1606, 0.1606, 1]. The structure of the ground state determined by PND is somewhat consistent with this, giving a magnetic moment  of [1.39(5),1.01(6),-4.18(2)]~$\mu_B$/Er$^{3+}$. \newline

The nature of the observed fractionalized magnetization plateaus is worth some discussion. Magnetization plateaus are not unheard of on the SSL. As mentioned previously, SrCu$_2$(BO$_3$)$_2$ exhibits fractionalized magnetization plateaus, which have been observed at many fractions of the full moment. Of particular relevance are fractions of $\frac{1}{4}$ and $\frac{1}{2}$ \cite{kageyama1999exact, matsuda2013magnetization,jaime2012magnetostriction,onizuka20001}, although many more are reported. These plateaus are due to the frustration-induced \cite{lacroix2011introduction} localization of the triplet wavefunctions, which form a regular square lattice \cite{miyahara1999exact}.

While the plateaus of SrCu$_2$(BO$_3$)$_2$ are quantum in nature (singlets have no classical analog), the $\frac{1}{4}$ magnetization plateau seen in Er$_2$Be$_2$GeO$_7$ can be described with a classical description, which is observed directly through PND. The deduced net moment of 1.39(5)~$\mu_B$/Er$^3$ is close to the observed plateau moment of 1.74~$\mu_B$/Er$^3$, indicating the saturation moment of the predicted magnetic space group agrees with the plateau moment. The arrangement of moments inside the 18.19 magnetic spacegroup shows some qualitative agreement with the predicted local z-axis of the PCCEF calculations. The moment of [1.39(5), 1.01(6), -4.18(2)] implies that the local direction is [0.33, 0.24, -1], compared to the predicted [-0.1606, 0.1606, 1] (note that the single-ion anistropy does not dictate which direction the moment should take, so [0.1606, -0.1606, -1] is an equally valid local z axis). The canting of the nearest-neighbour antiferromagnetic moments can be attributed to the single-ion anistropy, although other effects, such as the Dyzaloshinskii-Moriya interaction, may contribute as well. A change in the magnetic structure is implied past H$_C^2$, hence the canted antiferromagnetic structure cannot account for the observed hysteresis, which is typical of ferromagnetic structures. Naively, a fraction of $\frac{1}{2}$ would suggest a three-up-one-down-type structure. PND performed with a field between H$_C^3$ and H$_C^4$ could confirm this.

As mentioned above, the stabilization mechanism for $\frac{1}{4}$ and $\frac{1}{2}$ is known to be due to frustration in SrCu$_2$(BO$_3$)$_2$ (as well as the source of magnetization plateaus in general). Possibly, this may also be the source of the plateaus in Er$_2$Be$_2$GeO$_7$ as well, although theoretical investigation will be necessary to understand the origin of the plateaus beyond a phenomenological lens, and there is some evidence for stabilization via the Dyzaloshinskii-Moriya interaction \cite{yadav2024observation}. In addition, the isomorphic SSL material Er$_2$Be$_2$SiO$_7$ shows no evidence of hosting magnetization plateaus \cite{brassington2024magnetic}. The single-ion anisotropy is quasi-XY, while the zero-field magnetic structure is similar but lacks the canted nature seen in Er$_2$Be$_2$GeO$_7$. This implies that the nature of the magnetic properties are heavily influenced by the cation on the 2a Wyckoff position.

It should be noted that the critical fields observed in Er$_2$Be$_2$GeO$_7$ are significantly lower than most other SSL materials exhibiting magnetization plateaus. The lowest critical field in SrCu$_2$(BO$_3$)$_2$, for instance, is in excess of 20~T \cite{PhysRevLett.82.3168, onizuka20001}, and is inaccessible for study in conventional commercial magnetometers. The low critical fields observed in this SSL melilite system may prove extremely useful in future studies of fractionalized magnetization. Theoretical investigation into the origin of the magnetization plateaus in Er$_2$Be$_2$GeO$_7$ would prove beneficial. Study of the experimental CEFs and anisotropies could elicit the role of anisotropy on the magnetization plateaus.

\section{Conclusion}

We have shown the groundstate of Er$_2$Be$_2$GeO$_7$ is an Ising antiferromagnet with a magnetic structure (18.19) comprised of canted nearest-neighbour antiferromagnetic spins. We show that Er$_2$Be$_2$GeO$_7$ exhibits fractionalized magnetization plateaus at $\frac{1}{4}$ and $\frac{1}{2}$ of the full powder moment, and that the $\frac{1}{4}$ plateau is classically accounted for. When comparing Er$_2$Be$_2$GeO$_7$ with other known rare-earth melilites, like Yb$_2$Be$_2$GeO$_7$ \cite{pula2024candidate}, the qualities of each material vary substantially despite residing on isomorphic lattices, eluding to the tunability of the magnetic properties with the choice of the rare-earth site. In addition, the Wyckoff position 2a (Ge site) also appears to be important, as the properties of Er$_2$Be$_2$GeO$_7$ vary significantly from those reported on Er$_2$Be$_2$SiO$_7$ \cite{brassington2024magnetic}. Further experimental study of the rare-earth melilities and theoretical investigation into the origin of the magnetization plateaus seen in this work would likely prove most interesting.

\section{Acknowledgements}

We thank Dr. Bruce Gaulin for helpful discussions. Work at McMaster University was supported by the Natural Sciences and Engineering Research Council of Canada. A portion of this research used resources at the High Flux Isotope Reactor, a DOE Office of Science User Facility operated by the Oak Ridge National Laboratory. The beam time was allocated to HB-2A in proposal number IPTS-30630.1. Work at the University of Birmingham was supported by the UKRI Engineering and Physical Sciences Research Council (EP/V028774/1). The authors also acknowledge support from the University of Birmingham and McMaster University through the BIRMAC Quantum Materials Fund.


\begin{thebibliography}{25}%
\makeatletter
\providecommand \@ifxundefined [1]{%
 \@ifx{#1\undefined}
}%
\providecommand \@ifnum [1]{%
 \ifnum #1\expandafter \@firstoftwo
 \else \expandafter \@secondoftwo
 \fi
}%
\providecommand \@ifx [1]{%
 \ifx #1\expandafter \@firstoftwo
 \else \expandafter \@secondoftwo
 \fi
}%
\providecommand \natexlab [1]{#1}%
\providecommand \enquote  [1]{``#1''}%
\providecommand \bibnamefont  [1]{#1}%
\providecommand \bibfnamefont [1]{#1}%
\providecommand \citenamefont [1]{#1}%
\providecommand \href@noop [0]{\@secondoftwo}%
\providecommand \href [0]{\begingroup \@sanitize@url \@href}%
\providecommand \@href[1]{\@@startlink{#1}\@@href}%
\providecommand \@@href[1]{\endgroup#1\@@endlink}%
\providecommand \@sanitize@url [0]{\catcode `\\12\catcode `\$12\catcode
  `\&12\catcode `\#12\catcode `\^12\catcode `\_12\catcode `\%12\relax}%
\providecommand \@@startlink[1]{}%
\providecommand \@@endlink[0]{}%
\providecommand \url  [0]{\begingroup\@sanitize@url \@url }%
\providecommand \@url [1]{\endgroup\@href {#1}{\urlprefix }}%
\providecommand \urlprefix  [0]{URL }%
\providecommand \Eprint [0]{\href }%
\providecommand \doibase [0]{https://doi.org/}%
\providecommand \selectlanguage [0]{\@gobble}%
\providecommand \bibinfo  [0]{\@secondoftwo}%
\providecommand \bibfield  [0]{\@secondoftwo}%
\providecommand \translation [1]{[#1]}%
\providecommand \BibitemOpen [0]{}%
\providecommand \bibitemStop [0]{}%
\providecommand \bibitemNoStop [0]{.\EOS\space}%
\providecommand \EOS [0]{\spacefactor3000\relax}%
\providecommand \BibitemShut  [1]{\csname bibitem#1\endcsname}%
\let\auto@bib@innerbib\@empty
\bibitem [{\citenamefont {Rau}\ and\ \citenamefont
  {Gingras}(2019)}]{rau2019frustrated}%
  \BibitemOpen
  \bibfield  {author} {\bibinfo {author} {\bibfnamefont {J.~G.}\ \bibnamefont
  {Rau}}\ and\ \bibinfo {author} {\bibfnamefont {M.~J.}\ \bibnamefont
  {Gingras}},\ }\bibfield  {title} {\bibinfo {title} {Frustrated quantum
  rare-earth pyrochlores},\ }\href@noop {} {\bibfield  {journal} {\bibinfo
  {journal} {Annual Review of Condensed Matter Physics}\ }\textbf {\bibinfo
  {volume} {10}},\ \bibinfo {pages} {357} (\bibinfo {year} {2019})}\BibitemShut
  {NoStop}%
\bibitem [{\citenamefont {Lacroix}\ \emph {et~al.}(2011)\citenamefont
  {Lacroix}, \citenamefont {Mendels},\ and\ \citenamefont
  {Mila}}]{lacroix2011introduction}%
  \BibitemOpen
  \bibfield  {author} {\bibinfo {author} {\bibfnamefont {C.}~\bibnamefont
  {Lacroix}}, \bibinfo {author} {\bibfnamefont {P.}~\bibnamefont {Mendels}},\
  and\ \bibinfo {author} {\bibfnamefont {F.}~\bibnamefont {Mila}},\ }\href@noop
  {} {\emph {\bibinfo {title} {Introduction to frustrated magnetism: materials,
  experiments, theory}}},\ Vol.\ \bibinfo {volume} {164}\ (\bibinfo
  {publisher} {Springer Science \& Business Media},\ \bibinfo {year}
  {2011})\BibitemShut {NoStop}%
\bibitem [{\citenamefont {Balents}(2010)}]{balents2010spin}%
  \BibitemOpen
  \bibfield  {author} {\bibinfo {author} {\bibfnamefont {L.}~\bibnamefont
  {Balents}},\ }\bibfield  {title} {\bibinfo {title} {Spin liquids in
  frustrated magnets},\ }\href@noop {} {\bibfield  {journal} {\bibinfo
  {journal} {nature}\ }\textbf {\bibinfo {volume} {464}},\ \bibinfo {pages}
  {199} (\bibinfo {year} {2010})}\BibitemShut {NoStop}%
\bibitem [{\citenamefont {{Sriram Shastry}}\ and\ \citenamefont
  {Sutherland}(1981)}]{SRIRAMSHASTRY19811069}%
  \BibitemOpen
  \bibfield  {author} {\bibinfo {author} {\bibfnamefont {B.}~\bibnamefont
  {{Sriram Shastry}}}\ and\ \bibinfo {author} {\bibfnamefont {B.}~\bibnamefont
  {Sutherland}},\ }\bibfield  {title} {\bibinfo {title} {Exact ground state of
  a quantum mechanical antiferromagnet},\ }\href
  {https://doi.org/https://doi.org/10.1016/0378-4363(81)90838-X} {\bibfield
  {journal} {\bibinfo  {journal} {Physica B+C}\ }\textbf {\bibinfo {volume}
  {108}},\ \bibinfo {pages} {1069} (\bibinfo {year} {1981})}\BibitemShut
  {NoStop}%
\bibitem [{\citenamefont {Kageyama}\ \emph
  {et~al.}(1999{\natexlab{a}})\citenamefont {Kageyama}, \citenamefont
  {Yoshimura}, \citenamefont {Stern}, \citenamefont {Mushnikov}, \citenamefont
  {Onizuka}, \citenamefont {Kato}, \citenamefont {Kosuge}, \citenamefont
  {Slichter}, \citenamefont {Goto},\ and\ \citenamefont
  {Ueda}}]{PhysRevLett.82.3168}%
  \BibitemOpen
  \bibfield  {author} {\bibinfo {author} {\bibfnamefont {H.}~\bibnamefont
  {Kageyama}}, \bibinfo {author} {\bibfnamefont {K.}~\bibnamefont {Yoshimura}},
  \bibinfo {author} {\bibfnamefont {R.}~\bibnamefont {Stern}}, \bibinfo
  {author} {\bibfnamefont {N.~V.}\ \bibnamefont {Mushnikov}}, \bibinfo {author}
  {\bibfnamefont {K.}~\bibnamefont {Onizuka}}, \bibinfo {author} {\bibfnamefont
  {M.}~\bibnamefont {Kato}}, \bibinfo {author} {\bibfnamefont {K.}~\bibnamefont
  {Kosuge}}, \bibinfo {author} {\bibfnamefont {C.~P.}\ \bibnamefont
  {Slichter}}, \bibinfo {author} {\bibfnamefont {T.}~\bibnamefont {Goto}},\
  and\ \bibinfo {author} {\bibfnamefont {Y.}~\bibnamefont {Ueda}},\ }\bibfield
  {title} {\bibinfo {title} {Exact dimer ground state and quantized
  magnetization plateaus in the two-dimensional spin system
  ${\mathrm{srcu}}_{2}({\mathrm{bo}}_{3}){}_{2}$},\ }\href
  {https://doi.org/10.1103/PhysRevLett.82.3168} {\bibfield  {journal} {\bibinfo
   {journal} {Phys. Rev. Lett.}\ }\textbf {\bibinfo {volume} {82}},\ \bibinfo
  {pages} {3168} (\bibinfo {year} {1999}{\natexlab{a}})}\BibitemShut {NoStop}%
\bibitem [{\citenamefont {Koga}\ and\ \citenamefont
  {Kawakami}(2000)}]{koga2000quantum}%
  \BibitemOpen
  \bibfield  {author} {\bibinfo {author} {\bibfnamefont {A.}~\bibnamefont
  {Koga}}\ and\ \bibinfo {author} {\bibfnamefont {N.}~\bibnamefont
  {Kawakami}},\ }\bibfield  {title} {\bibinfo {title} {Quantum phase
  transitions in the shastry-sutherland model for {{SrCu$_2$(BO$_3$)$_2$}}},\
  }\href@noop {} {\bibfield  {journal} {\bibinfo  {journal} {Physical Review
  Letters}\ }\textbf {\bibinfo {volume} {84}},\ \bibinfo {pages} {4461}
  (\bibinfo {year} {2000})}\BibitemShut {NoStop}%
\bibitem [{\citenamefont {Guo}\ \emph {et~al.}(2020)\citenamefont {Guo},
  \citenamefont {Sun}, \citenamefont {Zhao}, \citenamefont {Wang},
  \citenamefont {Hong}, \citenamefont {Sidorov}, \citenamefont {Ma},
  \citenamefont {Wu}, \citenamefont {Li}, \citenamefont {Meng} \emph
  {et~al.}}]{guo2020quantum}%
  \BibitemOpen
  \bibfield  {author} {\bibinfo {author} {\bibfnamefont {J.}~\bibnamefont
  {Guo}}, \bibinfo {author} {\bibfnamefont {G.}~\bibnamefont {Sun}}, \bibinfo
  {author} {\bibfnamefont {B.}~\bibnamefont {Zhao}}, \bibinfo {author}
  {\bibfnamefont {L.}~\bibnamefont {Wang}}, \bibinfo {author} {\bibfnamefont
  {W.}~\bibnamefont {Hong}}, \bibinfo {author} {\bibfnamefont {V.~A.}\
  \bibnamefont {Sidorov}}, \bibinfo {author} {\bibfnamefont {N.}~\bibnamefont
  {Ma}}, \bibinfo {author} {\bibfnamefont {Q.}~\bibnamefont {Wu}}, \bibinfo
  {author} {\bibfnamefont {S.}~\bibnamefont {Li}}, \bibinfo {author}
  {\bibfnamefont {Z.~Y.}\ \bibnamefont {Meng}}, \emph {et~al.},\ }\bibfield
  {title} {\bibinfo {title} {Quantum phases of
  {${\mathrm{SrCu}}_{2}({\mathrm{BO}}_{3}){}_{2}$} from high-pressure
  thermodynamics},\ }\href@noop {} {\bibfield  {journal} {\bibinfo  {journal}
  {Physical Review Letters}\ }\textbf {\bibinfo {volume} {124}},\ \bibinfo
  {pages} {206602} (\bibinfo {year} {2020})}\BibitemShut {NoStop}%
\bibitem [{\citenamefont {Jim{\'e}nez}\ \emph {et~al.}(2021)\citenamefont
  {Jim{\'e}nez}, \citenamefont {Crone}, \citenamefont {Fogh}, \citenamefont
  {Zayed}, \citenamefont {Lortz}, \citenamefont {Pomjakushina}, \citenamefont
  {Conder}, \citenamefont {L{\"a}uchli}, \citenamefont {Weber}, \citenamefont
  {Wessel} \emph {et~al.}}]{jimenez2021quantum}%
  \BibitemOpen
  \bibfield  {author} {\bibinfo {author} {\bibfnamefont {J.~L.}\ \bibnamefont
  {Jim{\'e}nez}}, \bibinfo {author} {\bibfnamefont {S.}~\bibnamefont {Crone}},
  \bibinfo {author} {\bibfnamefont {E.}~\bibnamefont {Fogh}}, \bibinfo {author}
  {\bibfnamefont {M.~E.}\ \bibnamefont {Zayed}}, \bibinfo {author}
  {\bibfnamefont {R.}~\bibnamefont {Lortz}}, \bibinfo {author} {\bibfnamefont
  {E.}~\bibnamefont {Pomjakushina}}, \bibinfo {author} {\bibfnamefont
  {K.}~\bibnamefont {Conder}}, \bibinfo {author} {\bibfnamefont {A.~M.}\
  \bibnamefont {L{\"a}uchli}}, \bibinfo {author} {\bibfnamefont
  {L.}~\bibnamefont {Weber}}, \bibinfo {author} {\bibfnamefont
  {S.}~\bibnamefont {Wessel}}, \emph {et~al.},\ }\bibfield  {title} {\bibinfo
  {title} {A quantum magnetic analogue to the critical point of water},\
  }\href@noop {} {\bibfield  {journal} {\bibinfo  {journal} {Nature}\ }\textbf
  {\bibinfo {volume} {592}},\ \bibinfo {pages} {370} (\bibinfo {year}
  {2021})}\BibitemShut {NoStop}%
\bibitem [{\citenamefont {Wang}\ \emph {et~al.}(2022)\citenamefont {Wang},
  \citenamefont {Zhang},\ and\ \citenamefont {Sandvik}}]{Wang_2022}%
  \BibitemOpen
  \bibfield  {author} {\bibinfo {author} {\bibfnamefont {L.}~\bibnamefont
  {Wang}}, \bibinfo {author} {\bibfnamefont {Y.}~\bibnamefont {Zhang}},\ and\
  \bibinfo {author} {\bibfnamefont {A.~W.}\ \bibnamefont {Sandvik}},\
  }\bibfield  {title} {\bibinfo {title} {Quantum spin liquid phase in the
  {Shastry-Sutherland} model detected by an improved level spectroscopic
  method},\ }\href {https://doi.org/10.1088/0256-307X/39/7/077502} {\bibfield
  {journal} {\bibinfo  {journal} {Chinese Physics Letters}\ }\textbf {\bibinfo
  {volume} {39}},\ \bibinfo {pages} {077502} (\bibinfo {year}
  {2022})}\BibitemShut {NoStop}%
\bibitem [{\citenamefont {Kele\ifmmode~\mbox{\c{s}}\else \c{s}\fi{}}\ and\
  \citenamefont {Zhao}(2022)}]{PhysRevB.105.L041115}%
  \BibitemOpen
  \bibfield  {author} {\bibinfo {author} {\bibfnamefont {A.}~\bibnamefont
  {Kele\ifmmode~\mbox{\c{s}}\else \c{s}\fi{}}}\ and\ \bibinfo {author}
  {\bibfnamefont {E.}~\bibnamefont {Zhao}},\ }\bibfield  {title} {\bibinfo
  {title} {Rise and fall of plaquette order in the {Shastry-Sutherland} magnet
  revealed by pseudofermion functional renormalization group},\ }\href
  {https://doi.org/10.1103/PhysRevB.105.L041115} {\bibfield  {journal}
  {\bibinfo  {journal} {Phys. Rev. B}\ }\textbf {\bibinfo {volume} {105}},\
  \bibinfo {pages} {L041115} (\bibinfo {year} {2022})}\BibitemShut {NoStop}%
\bibitem [{\citenamefont {Yang}\ \emph {et~al.}(2022)\citenamefont {Yang},
  \citenamefont {Sandvik},\ and\ \citenamefont {Wang}}]{PhysRevB.105.L060409}%
  \BibitemOpen
  \bibfield  {author} {\bibinfo {author} {\bibfnamefont {J.}~\bibnamefont
  {Yang}}, \bibinfo {author} {\bibfnamefont {A.~W.}\ \bibnamefont {Sandvik}},\
  and\ \bibinfo {author} {\bibfnamefont {L.}~\bibnamefont {Wang}},\ }\bibfield
  {title} {\bibinfo {title} {Quantum criticality and spin liquid phase in the
  {Shastry-Sutherland model}},\ }\href
  {https://doi.org/10.1103/PhysRevB.105.L060409} {\bibfield  {journal}
  {\bibinfo  {journal} {Phys. Rev. B}\ }\textbf {\bibinfo {volume} {105}},\
  \bibinfo {pages} {L060409} (\bibinfo {year} {2022})}\BibitemShut {NoStop}%
\bibitem [{\citenamefont {Pula}\ \emph {et~al.}(2024)\citenamefont {Pula},
  \citenamefont {Sharma}, \citenamefont {Gautreau}, \citenamefont {KP},
  \citenamefont {Kanigel}, \citenamefont {Frontzek}, \citenamefont {Dolling},
  \citenamefont {Clark}, \citenamefont {Dunsiger}, \citenamefont {Ghara} \emph
  {et~al.}}]{pula2024candidate}%
  \BibitemOpen
  \bibfield  {author} {\bibinfo {author} {\bibfnamefont {M.}~\bibnamefont
  {Pula}}, \bibinfo {author} {\bibfnamefont {S.}~\bibnamefont {Sharma}},
  \bibinfo {author} {\bibfnamefont {J.}~\bibnamefont {Gautreau}}, \bibinfo
  {author} {\bibfnamefont {S.}~\bibnamefont {KP}}, \bibinfo {author}
  {\bibfnamefont {A.}~\bibnamefont {Kanigel}}, \bibinfo {author} {\bibfnamefont
  {M.}~\bibnamefont {Frontzek}}, \bibinfo {author} {\bibfnamefont
  {T.}~\bibnamefont {Dolling}}, \bibinfo {author} {\bibfnamefont
  {L.}~\bibnamefont {Clark}}, \bibinfo {author} {\bibfnamefont
  {S.}~\bibnamefont {Dunsiger}}, \bibinfo {author} {\bibfnamefont
  {A.}~\bibnamefont {Ghara}}, \emph {et~al.},\ }\bibfield  {title} {\bibinfo
  {title} {Candidate for a quantum spin liquid ground state in the
  {{Shastry-Sutherland}} lattice material {{Yb$_2$Be$_2$GeO$_7$}}},\
  }\href@noop {} {\bibfield  {journal} {\bibinfo  {journal} {Physical Review
  B}\ }\textbf {\bibinfo {volume} {110}},\ \bibinfo {pages} {014412} (\bibinfo
  {year} {2024})}\BibitemShut {NoStop}%
\bibitem [{\citenamefont {Ochi}\ \emph {et~al.}(1982)\citenamefont {Ochi},
  \citenamefont {Morikawa}, \citenamefont {Minato},\ and\ \citenamefont
  {Marumo}}]{OCHI1982911}%
  \BibitemOpen
  \bibfield  {author} {\bibinfo {author} {\bibfnamefont {Y.}~\bibnamefont
  {Ochi}}, \bibinfo {author} {\bibfnamefont {H.}~\bibnamefont {Morikawa}},
  \bibinfo {author} {\bibfnamefont {I.}~\bibnamefont {Minato}},\ and\ \bibinfo
  {author} {\bibfnamefont {F.}~\bibnamefont {Marumo}},\ }\bibfield  {title}
  {\bibinfo {title} {Preparation and magnetic property of new rare earth
  compounds {{R$_2$GeBe$_2$O$_7$ (R = La, Pr, Sm, Gd, Dy, Er) and
  Y$_2$GeBe$_2$O$_7$}}},\ }\href
  {https://doi.org/https://doi.org/10.1016/0025-5408(82)90012-5} {\bibfield
  {journal} {\bibinfo  {journal} {Materials Research Bulletin}\ }\textbf
  {\bibinfo {volume} {17}},\ \bibinfo {pages} {911} (\bibinfo {year}
  {1982})}\BibitemShut {NoStop}%
\bibitem [{\citenamefont {Ashtar}\ \emph {et~al.}(2021)\citenamefont {Ashtar},
  \citenamefont {Bai}, \citenamefont {Xu}, \citenamefont {Wan}, \citenamefont
  {Wei}, \citenamefont {Liu}, \citenamefont {Marwat},\ and\ \citenamefont
  {Tian}}]{doi:10.1021/acs.inorgchem.0c03131}%
  \BibitemOpen
  \bibfield  {author} {\bibinfo {author} {\bibfnamefont {M.}~\bibnamefont
  {Ashtar}}, \bibinfo {author} {\bibfnamefont {Y.}~\bibnamefont {Bai}},
  \bibinfo {author} {\bibfnamefont {L.}~\bibnamefont {Xu}}, \bibinfo {author}
  {\bibfnamefont {Z.}~\bibnamefont {Wan}}, \bibinfo {author} {\bibfnamefont
  {Z.}~\bibnamefont {Wei}}, \bibinfo {author} {\bibfnamefont {Y.}~\bibnamefont
  {Liu}}, \bibinfo {author} {\bibfnamefont {M.~A.}\ \bibnamefont {Marwat}},\
  and\ \bibinfo {author} {\bibfnamefont {Z.}~\bibnamefont {Tian}},\ }\bibfield
  {title} {\bibinfo {title} {Structure and magnetic properties of melilite-type
  compounds {{RE$_2$Be$_2$GeO$_7$ (RE = Pr, Nd, Gd–Yb)}} with rare-earth ions
  on {{Shastry–Sutherland}} lattice},\ }\href
  {https://doi.org/10.1021/acs.inorgchem.0c03131} {\bibfield  {journal}
  {\bibinfo  {journal} {Inorganic Chemistry}\ }\textbf {\bibinfo {volume}
  {60}},\ \bibinfo {pages} {3626} (\bibinfo {year} {2021})},\ \bibinfo {note}
  {pMID: 33635649},\ \Eprint
  {https://arxiv.org/abs/https://doi.org/10.1021/acs.inorgchem.0c03131}
  {https://doi.org/10.1021/acs.inorgchem.0c03131} \BibitemShut {NoStop}%
\bibitem [{\citenamefont {Brassington}\ \emph {et~al.}(2024)\citenamefont
  {Brassington}, \citenamefont {Ma}, \citenamefont {Sala}, \citenamefont
  {Kolesnikov}, \citenamefont {Taddei}, \citenamefont {Wu}, \citenamefont
  {Choi}, \citenamefont {Wang}, \citenamefont {Xie}, \citenamefont {Ma} \emph
  {et~al.}}]{brassington2024magnetic}%
  \BibitemOpen
  \bibfield  {author} {\bibinfo {author} {\bibfnamefont {A.}~\bibnamefont
  {Brassington}}, \bibinfo {author} {\bibfnamefont {Q.}~\bibnamefont {Ma}},
  \bibinfo {author} {\bibfnamefont {G.}~\bibnamefont {Sala}}, \bibinfo {author}
  {\bibfnamefont {A.}~\bibnamefont {Kolesnikov}}, \bibinfo {author}
  {\bibfnamefont {K.}~\bibnamefont {Taddei}}, \bibinfo {author} {\bibfnamefont
  {Y.}~\bibnamefont {Wu}}, \bibinfo {author} {\bibfnamefont {E.}~\bibnamefont
  {Choi}}, \bibinfo {author} {\bibfnamefont {H.}~\bibnamefont {Wang}}, \bibinfo
  {author} {\bibfnamefont {W.}~\bibnamefont {Xie}}, \bibinfo {author}
  {\bibfnamefont {J.}~\bibnamefont {Ma}}, \emph {et~al.},\ }\bibfield  {title}
  {\bibinfo {title} {Magnetic properties of the quasi-xy {{Shastry-Sutherland}}
  magnet {{Er$_2$Be$_2$SiO$_7$}}},\ }\href@noop {} {\bibfield  {journal}
  {\bibinfo  {journal} {Physical Review Materials}\ }\textbf {\bibinfo {volume}
  {8}},\ \bibinfo {pages} {094001} (\bibinfo {year} {2024})}\BibitemShut
  {NoStop}%
\bibitem [{\citenamefont {Momma}\ and\ \citenamefont
  {Izumi}(2011)}]{momma2011vesta}%
  \BibitemOpen
  \bibfield  {author} {\bibinfo {author} {\bibfnamefont {K.}~\bibnamefont
  {Momma}}\ and\ \bibinfo {author} {\bibfnamefont {F.}~\bibnamefont {Izumi}},\
  }\bibfield  {title} {\bibinfo {title} {Vesta 3 for three-dimensional
  visualization of crystal, volumetric and morphology data},\ }\href@noop {}
  {\bibfield  {journal} {\bibinfo  {journal} {Journal of applied
  crystallography}\ }\textbf {\bibinfo {volume} {44}},\ \bibinfo {pages} {1272}
  (\bibinfo {year} {2011})}\BibitemShut {NoStop}%
\bibitem [{\citenamefont {Scheie}(2021)}]{scheie2021pycrystalfield}%
  \BibitemOpen
  \bibfield  {author} {\bibinfo {author} {\bibfnamefont {A.}~\bibnamefont
  {Scheie}},\ }\bibfield  {title} {\bibinfo {title} {Pycrystalfield: software
  for calculation, analysis and fitting of crystal electric field
  hamiltonians},\ }\href@noop {} {\bibfield  {journal} {\bibinfo  {journal}
  {Journal of Applied Crystallography}\ }\textbf {\bibinfo {volume} {54}},\
  \bibinfo {pages} {356} (\bibinfo {year} {2021})}\BibitemShut {NoStop}%
\bibitem [{\citenamefont {Baur}(1956)}]{baur1956verfeinerung}%
  \BibitemOpen
  \bibfield  {author} {\bibinfo {author} {\bibfnamefont {W.~H.}\ \bibnamefont
  {Baur}},\ }\bibfield  {title} {\bibinfo {title} {{\"U}ber die verfeinerung
  der kristallstrukturbestimmung einiger vertreter des rutiltyps: {{TiO$_2$,
  SnO$_2$, GeO$_2$ und MgF$_2$}}},\ }\href@noop {} {\bibfield  {journal}
  {\bibinfo  {journal} {Acta Crystallographica}\ }\textbf {\bibinfo {volume}
  {9}},\ \bibinfo {pages} {515} (\bibinfo {year} {1956})}\BibitemShut {NoStop}%
\bibitem [{\citenamefont {Jain}\ \emph {et~al.}(2007)\citenamefont {Jain},
  \citenamefont {Gupta}, \citenamefont {Gupta},\ and\ \citenamefont
  {Kumar}}]{LandoltBornstein2007}%
  \BibitemOpen
  \bibfield  {author} {\bibinfo {author} {\bibfnamefont {M.}~\bibnamefont
  {Jain}}, \bibinfo {author} {\bibfnamefont {A.}~\bibnamefont {Gupta}},
  \bibinfo {author} {\bibfnamefont {R.}~\bibnamefont {Gupta}},\ and\ \bibinfo
  {author} {\bibfnamefont {M.}~\bibnamefont {Kumar}},\ }\href@noop {} {\emph
  {\bibinfo {title} {Diamagnetic Susceptibility and Anisotropy of Inorganic and
  Organometallic Compounds}}},\ Vol.~\bibinfo {volume} {27}\ (\bibinfo
  {publisher} {Springer Science \& Business Media},\ \bibinfo {year}
  {2007})\BibitemShut {NoStop}%
\bibitem [{\citenamefont {Kageyama}\ \emph
  {et~al.}(1999{\natexlab{b}})\citenamefont {Kageyama}, \citenamefont
  {Yoshimura}, \citenamefont {Stern}, \citenamefont {Mushnikov}, \citenamefont
  {Onizuka}, \citenamefont {Kato}, \citenamefont {Kosuge}, \citenamefont
  {Slichter}, \citenamefont {Goto},\ and\ \citenamefont
  {Ueda}}]{kageyama1999exact}%
  \BibitemOpen
  \bibfield  {author} {\bibinfo {author} {\bibfnamefont {H.}~\bibnamefont
  {Kageyama}}, \bibinfo {author} {\bibfnamefont {K.}~\bibnamefont {Yoshimura}},
  \bibinfo {author} {\bibfnamefont {R.}~\bibnamefont {Stern}}, \bibinfo
  {author} {\bibfnamefont {N.}~\bibnamefont {Mushnikov}}, \bibinfo {author}
  {\bibfnamefont {K.}~\bibnamefont {Onizuka}}, \bibinfo {author} {\bibfnamefont
  {M.}~\bibnamefont {Kato}}, \bibinfo {author} {\bibfnamefont {K.}~\bibnamefont
  {Kosuge}}, \bibinfo {author} {\bibfnamefont {C.}~\bibnamefont {Slichter}},
  \bibinfo {author} {\bibfnamefont {T.}~\bibnamefont {Goto}},\ and\ \bibinfo
  {author} {\bibfnamefont {Y.}~\bibnamefont {Ueda}},\ }\bibfield  {title}
  {\bibinfo {title} {Exact dimer ground state and quantized magnetization
  plateaus in the two-dimensional spin system {{SrCu$_2$(BO$_3$)$_2$}}},\
  }\href@noop {} {\bibfield  {journal} {\bibinfo  {journal} {Physical review
  letters}\ }\textbf {\bibinfo {volume} {82}},\ \bibinfo {pages} {3168}
  (\bibinfo {year} {1999}{\natexlab{b}})}\BibitemShut {NoStop}%
\bibitem [{\citenamefont {Matsuda}\ \emph {et~al.}(2013)\citenamefont
  {Matsuda}, \citenamefont {Abe}, \citenamefont {Takeyama}, \citenamefont
  {Kageyama}, \citenamefont {Corboz}, \citenamefont {Honecker}, \citenamefont
  {Manmana}, \citenamefont {Foltin}, \citenamefont {Schmidt},\ and\
  \citenamefont {Mila}}]{matsuda2013magnetization}%
  \BibitemOpen
  \bibfield  {author} {\bibinfo {author} {\bibfnamefont {Y.~H.}\ \bibnamefont
  {Matsuda}}, \bibinfo {author} {\bibfnamefont {N.}~\bibnamefont {Abe}},
  \bibinfo {author} {\bibfnamefont {S.}~\bibnamefont {Takeyama}}, \bibinfo
  {author} {\bibfnamefont {H.}~\bibnamefont {Kageyama}}, \bibinfo {author}
  {\bibfnamefont {P.}~\bibnamefont {Corboz}}, \bibinfo {author} {\bibfnamefont
  {A.}~\bibnamefont {Honecker}}, \bibinfo {author} {\bibfnamefont {S.~R.}\
  \bibnamefont {Manmana}}, \bibinfo {author} {\bibfnamefont {G.}~\bibnamefont
  {Foltin}}, \bibinfo {author} {\bibfnamefont {K.}~\bibnamefont {Schmidt}},\
  and\ \bibinfo {author} {\bibfnamefont {F.}~\bibnamefont {Mila}},\ }\bibfield
  {title} {\bibinfo {title} {Magnetization of {{SrCu$_2$(BO$_3$)$_2$}} in
  ultrahigh magnetic fields up to 118 {{T}}},\ }\href@noop {} {\bibfield
  {journal} {\bibinfo  {journal} {Physical review letters}\ }\textbf {\bibinfo
  {volume} {111}},\ \bibinfo {pages} {137204} (\bibinfo {year}
  {2013})}\BibitemShut {NoStop}%
\bibitem [{\citenamefont {Jaime}\ \emph {et~al.}(2012)\citenamefont {Jaime},
  \citenamefont {Daou}, \citenamefont {Crooker}, \citenamefont {Weickert},
  \citenamefont {Uchida}, \citenamefont {Feiguin}, \citenamefont {Batista},
  \citenamefont {Dabkowska},\ and\ \citenamefont
  {Gaulin}}]{jaime2012magnetostriction}%
  \BibitemOpen
  \bibfield  {author} {\bibinfo {author} {\bibfnamefont {M.}~\bibnamefont
  {Jaime}}, \bibinfo {author} {\bibfnamefont {R.}~\bibnamefont {Daou}},
  \bibinfo {author} {\bibfnamefont {S.~A.}\ \bibnamefont {Crooker}}, \bibinfo
  {author} {\bibfnamefont {F.}~\bibnamefont {Weickert}}, \bibinfo {author}
  {\bibfnamefont {A.}~\bibnamefont {Uchida}}, \bibinfo {author} {\bibfnamefont
  {A.~E.}\ \bibnamefont {Feiguin}}, \bibinfo {author} {\bibfnamefont {C.~D.}\
  \bibnamefont {Batista}}, \bibinfo {author} {\bibfnamefont {H.~A.}\
  \bibnamefont {Dabkowska}},\ and\ \bibinfo {author} {\bibfnamefont {B.~D.}\
  \bibnamefont {Gaulin}},\ }\bibfield  {title} {\bibinfo {title}
  {Magnetostriction and magnetic texture to 100.75 {{Tesla}} in frustrated
  {{SrCu$_2$(BO$_3$)$_2$}}},\ }\href@noop {} {\bibfield  {journal} {\bibinfo
  {journal} {Proceedings of the National Academy of Sciences}\ }\textbf
  {\bibinfo {volume} {109}},\ \bibinfo {pages} {12404} (\bibinfo {year}
  {2012})}\BibitemShut {NoStop}%
\bibitem [{\citenamefont {Onizuka}\ \emph {et~al.}(2000)\citenamefont
  {Onizuka}, \citenamefont {Kageyama}, \citenamefont {Narumi}, \citenamefont
  {Kindo}, \citenamefont {Ueda},\ and\ \citenamefont {Goto}}]{onizuka20001}%
  \BibitemOpen
  \bibfield  {author} {\bibinfo {author} {\bibfnamefont {K.}~\bibnamefont
  {Onizuka}}, \bibinfo {author} {\bibfnamefont {H.}~\bibnamefont {Kageyama}},
  \bibinfo {author} {\bibfnamefont {Y.}~\bibnamefont {Narumi}}, \bibinfo
  {author} {\bibfnamefont {K.}~\bibnamefont {Kindo}}, \bibinfo {author}
  {\bibfnamefont {Y.}~\bibnamefont {Ueda}},\ and\ \bibinfo {author}
  {\bibfnamefont {T.}~\bibnamefont {Goto}},\ }\bibfield  {title} {\bibinfo
  {title} {$\frac{1}{3}$ magnetization plateau in
  {{SrCu$_2$(BO$_3$)$_2$}}-stripe order of excited triplets},\ }\href@noop {}
  {\bibfield  {journal} {\bibinfo  {journal} {Journal of the Physical Society
  of Japan}\ }\textbf {\bibinfo {volume} {69}},\ \bibinfo {pages} {1016}
  (\bibinfo {year} {2000})}\BibitemShut {NoStop}%
\bibitem [{\citenamefont {Miyahara}\ and\ \citenamefont
  {Ueda}(1999)}]{miyahara1999exact}%
  \BibitemOpen
  \bibfield  {author} {\bibinfo {author} {\bibfnamefont {S.}~\bibnamefont
  {Miyahara}}\ and\ \bibinfo {author} {\bibfnamefont {K.}~\bibnamefont
  {Ueda}},\ }\bibfield  {title} {\bibinfo {title} {Exact dimer ground state of
  the two dimensional heisenberg spin system {{SrCu$_2$(BO$_3$)$_2$}}},\
  }\href@noop {} {\bibfield  {journal} {\bibinfo  {journal} {Physical review
  letters}\ }\textbf {\bibinfo {volume} {82}},\ \bibinfo {pages} {3701}
  (\bibinfo {year} {1999})}\BibitemShut {NoStop}%
\bibitem [{\citenamefont {Yadav}\ \emph {et~al.}(2024)\citenamefont {Yadav},
  \citenamefont {Rufino}, \citenamefont {Bag}, \citenamefont {Kolesnikov},
  \citenamefont {Garlea}, \citenamefont {Graf}, \citenamefont {Mila},
  \citenamefont {Haravifard} \emph {et~al.}}]{yadav2024observation}%
  \BibitemOpen
  \bibfield  {author} {\bibinfo {author} {\bibfnamefont {L.}~\bibnamefont
  {Yadav}}, \bibinfo {author} {\bibfnamefont {A.}~\bibnamefont {Rufino}},
  \bibinfo {author} {\bibfnamefont {R.}~\bibnamefont {Bag}}, \bibinfo {author}
  {\bibfnamefont {A.~I.}\ \bibnamefont {Kolesnikov}}, \bibinfo {author}
  {\bibfnamefont {V.~O.}\ \bibnamefont {Garlea}}, \bibinfo {author}
  {\bibfnamefont {D.}~\bibnamefont {Graf}}, \bibinfo {author} {\bibfnamefont
  {F.}~\bibnamefont {Mila}}, \bibinfo {author} {\bibfnamefont {S.}~\bibnamefont
  {Haravifard}}, \emph {et~al.},\ }\bibfield  {title} {\bibinfo {title}
  {Observation of unprecedented fractional magnetization plateaus in a new
  {{Shastry-Sutherland}} ising compound},\ }\href@noop {} {\bibfield  {journal}
  {\bibinfo  {journal} {arXiv preprint arXiv:2405.12405}\ } (\bibinfo {year}
  {2024})}\BibitemShut {NoStop}%
\end{thebibliography}
\end{document}